\documentstyle[epsf,pra,aps]{revtex}

\newcommand{\cA}{{\cal A}}

\newcommand{\cC}{{\cal C}}
\newcommand{\cD}{{\cal D}}

\newcommand{\cG}{{\cal G}}
\newcommand{\cH}{{\cal H}}
\newcommand{\cJ}{{\cal J}}
\newcommand{\cL}{{\cal L}}
\newcommand{\cN}{{\cal N}}

\newcommand{\cR}{{\cal R}}
\newcommand{\cS}{{\cal S}}
\newcommand{\cU}{{\cal U}}
\newcommand{\cZ}{{\cal Z}}

\begin{document}
\preprint{}
\draft

\title{ On Quantum Control via Encoded Dynamical Decoupling } 
 
\author{ Lorenza Viola\thanks{Electronic address: lviola@lanl.gov }}
\address{ Los Alamos National Laboratory, Los Alamos, New Mexico 87545, 
USA} 

\maketitle

\begin{abstract}
I revisit the ideas underlying dynamical decoupling methods within the 
framework of quantum information processing, and examine their potential 
for direct implementations in terms of encoded rather than physical 
degrees of freedom. The usefulness of encoded decoupling schemes as a 
tool for engineering both closed- and open-system encoded evolutions
is investigated based on simple examples.
\end{abstract}



\section{Introduction}

Since the pioneering work on coherent averaging effects  
by Haeberlen and Waugh \cite{haeberlen0}, the use of tailored pulse 
sequences for manipulating the effective Hamiltonian experienced by a 
target quantum system has developed a solid tradition in nuclear 
magnetic resonance (NMR) \cite{haeberlen,ernst}. 
In particular, within the context of NMR 
quantum information processing (QIP), decoupling and refocusing 
techniques provide the basic tools for enforcing Hamiltonian 
evolutions that correspond to quantum logic gates between selected 
spins \cite{cory}.
The principles underlying these techniques, along with the powerful
formalism offered by average Hamiltonian theory (AHT)
\cite{haeberlen0}, have been recently extended beyond the NMR domain, 
and suggestive applications have resulted in various directions 
within QIP. 
On one hand, ideas from NMR decoupling motivated a ``bang-bang'' 
\cite{pra} control-theoretic framework for generic open quantum systems 
\cite{prl1}, which paved the way for the development of quantum 
symmetrization procedures and quantum error suppression strategies 
for QIP \cite{prl1,zanardi1,prl2,dygen,tombesi}. 
On the other hand, the application of active dynamical control in the
bang-bang limit proved a valuable tool for engineering the evolution 
of coupled quantum subsystems \cite{prl2}, leading to various 
schemes for universal simulations of both closed-system 
\cite{beth} and open-system dynamics \cite{lloyd}. 

So far, in spite of the pervasive role played by quantum coding in 
QIP, the application of active refocusing and decoupling methods
has been primarily thought of in terms of the basic {\sl physical} 
degrees of freedom. Two exceptions are a proposal by Wu and Lidar for 
applying recoupling schemes on encoded qubits governed by exchange-type 
Hamiltonians \cite{wu}, and an implementation by Fortunato {\it et al.} 
of encoded refocusing to gain universal control on a decoherence-free 
qubit \cite{dfs}. 
It is the purpose of this note to further comment on the significance 
of dynamical control methods as directly represented in terms of
{\sl encoded} degrees of freedom, by continuing the investigation 
of the interplay between quantum coding and decoupling techniques 
undertaken in \cite{dygen}, and expanding the basic arguments sketched 
in \cite{dfs}. While decoupling methods have been already shown to 
enable, in principle, to synthesize effective evolutions supporting 
noise-protected, encoded degrees of freedom \cite{dygen,zanardi2}, 
a different perspective is taken here -- by imagining a pre-selected
encoded structure, and by looking at the evolutions that can be 
enforced through encoded decoupling sequences. 

I envision two prospective types of applications of encoded 
dynamical decoupling methods within QIP. Similar to their 
un-encoded counterparts, these include control of 
both closed-system Hamiltonian evolutions and open-system noisy
evolutions -- all evolutions being however restricted to an 
underlying coding space that is preserved throughout. After
summarizing the basic facts about decoupling in Sect. II, 
encoded dynamical decoupling is introduced in Sect. III.
I then discuss the two relevant areas of application based on 
simple representative examples in Sect. IV and V, respectively. 
A brief summary concludes in Sect. VI.

\section{Basics of dynamical decoupling}

A bang-bang (b.b.) quantum control problem is concerned with 
characterizing the effective evolutions that can be engineered
by repeatedly interspersing, according to various possible schemes, 
the natural dynamics of a quantum system with full-power, 
instantaneous control operations (b.b. controls) 
\cite{prl1,zanardi1,prl2,byrd}. 
Let $S$ denote the target control system, defined on a 
(finite-dimensional) state space ${\cal H}_S$, dim(${\cH}_S)=N$,
$N=2^n$ for qubit systems. In a general open-system setting, 
$S$ is coupled to an environment $E$ via an interaction Hamiltonian
$H_{SE}$. The control problem can then be formulated in terms of the 
following data: \\
\noindent 
$\bullet$ $H$, the natural Hamiltonian of the joint 
system, $H=H_S \otimes \openone_E + \openone_S \otimes H_E +
H_{SE}$, determining the free unitary evolution $U_0(t)=
\exp(-iHt)$ on the joint state space ${\cal H}_S \otimes 
{\cal H}_E$. One can write $H_{SE}=\sum_\alpha S_\alpha \otimes 
E_\alpha$, for appropriate system ($S_\alpha$) and environment 
($E_\alpha$) operators. The linear subspace ${\cal N}=
\text{span}\{S_\alpha\}$ of noise-inducing couplings is referred to 
as {\sl error space}. Without loss of generality, both $H_S$ and the 
error generators in ${\cal N}$ can be assumed to be traceless. Detailed 
knowledge of $H_S$ or ${\cal N}$ may or may be not explicitly 
available from the start. \\
\noindent 
$\bullet$ ${\cal G}_{b.b.}$, the set of {\sl all} b.b. operations that 
can be effected on $S$. As it is conceivable that $P^{-1}=P^\dagger
\in {\cal G}_{b.b.}$ if $P \in {\cal G}_{b.b.}$, ${\cal G}$ 
can be associated with a subgroup of the group 
${\cal U}({\cal H})$ of unitary transformations on the system alone; \\
\noindent 
$\bullet$ ${\cal G}$, the {\sl discrete subset} of b.b. decoupling 
operations, ${\cal G}= \{ U_k \} \subseteq {\cal G}_{b.b.}$, with 
$k \in {\cal K}$ for some finite set of indexes with order 
$|{\cal K}|\equiv|{\cal G}|$; \\
\noindent
$\bullet$ $T_c$, the relevant control time scale ({\sl cycle time}), 
associated with the duration of a single control cycle; \\
\noindent 
$\bullet$ $\{ \tau_k\}$, the set of relative temporal separations between 
consecutive b.b. operations, $\tau_k =\Delta t_k/T_c >0$, 
in terms of the free evolution intervals $\Delta t_k$, and 
$\sum_{k \in {\cal K}}\tau_k =1$.  

AHT provides a general prescription for characterizing the controlled 
evolutions in terms of time-independent effective Hamiltonians that 
would result in the same unitary propagator if applied over the same 
evolution interval. Imagine that a single
control period $T_c$ consists of a cyclic sequence of $|{\cal K}|$ 
b.b. pulses, ${\cal P} =\{ P_k, \tau_k \}_{k=1}^{|{\cal K}|}$, 
with $\prod_{k=1}^{|{\cal K}|}P_k=\openone$. Then 
\begin{equation}
U(T_c)=\exp(-i H_{eff} T_c)= \prod_{m=0}^{M} 
U_m^\dagger U_0(\Delta t_m) U_m = \prod_{m=0}^{M} 
\exp(-i H_m \tau_m T_c) \:, 
\label{heff}
\end{equation}
where the first equality defines the effective average Hamiltonian, 
and the ``toggling-frame'' transformed Hamiltonians $H_m$ are 
determined by the composite rotations $U_m=\prod_{k=1}^m P_k$, 
$k=1, \ldots, M-1$, $U_0=\openone$. Here, $M=|{\cal K}|$ or 
$M=|{\cal K}|+1$ depending on whether the sequence is arranged so 
as to allow evolution in the $\openone$-frame in a single or a pair 
of control sub-intervals -- being, in any case, $U_{|{\cal K}|}= 
\openone$ by cyclicity. Because $P_k \in {\cal G}_{b.b.}$, 
then ${\cal G}\subseteq {\cG}_{b.b.}$ as anticipated. 
However, ${\cal G}$ need not itself be a subgroup, 
neither does ${\cal G}={\cG}_{b.b.}$ in general. For instance, 
allowing for ${\cal G}\not = {\cG}_{b.b.}$ may be crucial for
retaining universal control over decoupled dynamics \cite{prl2}.
Although the effective Hamiltonian (\ref{heff}) can be 
systematically calculated as a power series in the controllable 
parameter $T_c$ (Magnus series \cite{ernst}), 
AHT is practically useful in the limit of {\sl fast 
control} $T_c \rightarrow 0$ \cite{haeberlen,ernst,prl1}, where 
lowest-order contributions in $T_c$ suffice for an accurate description. 
In this limit, $H_{eff}$ approaches  
\begin{equation}
H_{eff} \mapsto \overline{H}= 
\sum_{k \in {\cal K}} \tau_k U_k^\dagger H U_k = \Lambda_{\cG}(H)\:, 
\label{qop}
\end{equation}
with leading corrections accounted for by 
\begin{equation}
\overline{H}^{(1)}= -{ i \over 2T_c \hbar} \sum_{m>n}\, [H_m,H_n]\,
\tau_m \tau_n \:. 
\label{first}
\end{equation} 
Under the conditions ensuring convergence of the series defining $H_{eff}$, 
this correction is at least $O(T_c)$. 
The zero-th order approximation (\ref{qop}) is, of course, exact if 
the various transformed Hamiltonians commute -- in which case any 
time scale constraint disappears. 

In passing, it is worth noting that the most general transformation 
that AHT allows involves, as given in (\ref{qop}), a weighted (convex)
mixture of unitary operators. Extended to the space of linear operators 
End$({\cH}_S)$ over $\cH_S$, the action (\ref{qop}) defines a 
trace-preserving, unital, completely positive map $\Lambda_{\cal G}$. 
Examples abound in NMR 
where decoupling is achieved by control actions of this form. A 
representative case is the so-called {\sc whh-4} sequence used for 
homo-nuclear dipolar decoupling \cite{haeberlen}. This corresponds to 
a sequence ${\cal P}=\{P_x, P_{-y}, P_y, P_{-x} \}$ of b.b. 
$\pi/2$-pulses, $P_a=\exp(-i \pi \sigma_a/4)$, and 
$\tau_0=\tau_1=\tau_3=\tau_4=1/6$, $\tau_2=1/3$ -- which effectively 
averages out two-spin interactions proportional to 
$3\,\sigma_z^i \sigma_z^j - \vec{\sigma}^i \cdot \vec{\sigma}^j$ 
over a cycle. 

The simplest realization of (\ref{qop}) occurs under the additional 
assumptions that control operations are equally separated in time and 
the composite rotations in ${\cal G}$ close a group ({\sl decoupling group} 
\cite{prl1}). By letting $\tau_k= 1/|{\cal G}|$, the quantum operation 
$\Lambda_{\cG}$ of (\ref{qop}) reduces in this case to the projector 
$\Pi_{\cal G}$ on the centralizer 
${\cal Z}({\cal G})$ of ${\cal G}$ in End$({\cH}_S)$,
${\cal Z}({\cal G})=\{ X \in \text{End}({\cH}_S)\,|\,
[X, U_k]=0 \;\forall k \}$ \cite{prl1,zanardi1,prl2}:
\begin{equation}
H_{eff} \mapsto \overline{H}= 
{1 \over |{\cal G}|}
\sum_{k \in {\cal K}} U_k^\dagger H U_k = \Pi_{\cG} ({H})\:. 
\label{projector}
\end{equation}
The resulting effective Hamiltonian acquires now a direct symmetry 
characterization as $[ \overline{H}, {\cG}]=0$ -- meaning that the 
controlled dynamics is symmetrized according to $\cG$ \cite{zanardi1},
and the desired averaging action can be thought of as filtering out 
the dynamics that is not invariant under $\cG$. 

One of the most famous (and practically useful) examples of a 
decoupling sequence with the above form is the Carr-Purcell ({\sc cp}) 
sequence \cite{carr,haeberlen,ernst}, which in 
its basic variant is used for suppressing undesired phase evolution 
due to $\sigma^z$ terms -- representing, for instance, applied field 
inhomogeneity in NMR. 
The sequence consists of repeated b.b. $\pi_a$-pulses, $a=x$ or $y$, 
separated by $\Delta t$, corresponding to a decoupling group 
${\cal G}_{{\sc cp}_a}=\{\openone, \pi_a \}$, 
with $\pi_a=\exp(-i\pi\sigma_a/2)$.  Within the cycle time
$T_c=2\Delta t$, the pulses can be arranged so as either 
$\tau_1=\tau_2=1/2$ (with $M=|{\cal K}|=2$) or $\tau_1=\tau_3=1/4$,
$\tau_2=1/2$ (with $M=|{\cal K}|+1=3$). 
Although the two sequences clearly lead to
the same $\overline{H}$ as resulting from (\ref{projector}), the second
option is actually superior in terms of the overall averaging accuracy
-- as it corresponds to a time-symmetric cycle for which all odd-order
corrections $\overline{H}^{(2\ell+1)}$, $\ell=0,1,\ldots$, vanish 
\cite{ernst}.

In a closed-system setting, $H_{SE}=0$ and the control problem consists 
in turning off (selectively or not) undesired contributions to $H_S$.
Note that this amounts to simulating the reachable AHT Hamiltonians
by $H_S$. In particular, b.b. control may enable universal 
Hamiltonian simulation if an arbitrary effective evolution can be
simulated by $H_S$. 
In an open-system setting, the focus is on averaging out the error
generators $\{ S_\alpha\}$, {\it i.e.} ensuring that $\Lambda_{\cG}
({\cal N})=0$, in order to achieve decoherence suppression.
The limit of fast control may be especially stringent to met in this
case, as it requires that $T_c \lesssim \tau_c$ for the shortest 
correlation time associated with the environmental noise 
\cite{pra,prl1,tombesi}
-- which can be prohibitively small in Markovian or quasi-Markovian 
dynamics. In both situations, the attainable control goals are 
influenced by two main factors: the available knowledge of the 
interactions to be manipulated and/or averaged out, and the overall 
available control resources.

Some notable results about decoupling are worth mentioning. \\
\noindent
$\bullet$ {\sl Selective averaging} (or refocusing) requires knowledge of
the transformation properties of the interactions to be turned off 
with respect to the achievable decoupling sets ${\cal G}$. 
For a decoupling group $\cG$, a necessary condition for selectivity is 
that the centralizer is non-trivial, ${\cal Z}(\cG) 
\not = \{ \lambda \openone \}$ ($\lambda$ complex) \cite{prl1,prl2}. 
A variety of {\sl efficient} schemes ({\it i.e.}, polynomial in $n$) 
exist for selective decoupling in systems with at most bilinear 
interactions \cite{mahler}. These are applicable to NMR and NMR-like 
Hamiltonians in both the weak and strong coupling limit. \\
\noindent
$\bullet$ For arbitrary interactions ($H_S$ or ${\cal N}$), {\sl maximal
averaging} (or complete decoupling, or annihilation) is possible in 
principle by letting ${\cal G}$ to be a unitary error basis on $\cH_S$
\cite{prl1,beth}.  Although this choice can be shown to be optimal, 
the resulting scheme is inefficient as the number of required b.b. 
control operations is $|\cG|=N^2=4^n$ which grows exponentially with
$n$. Again, efficient schemes can be designed if no terms higher than 
bilinear ones are known to be relevant \cite{mahler}.\\
\noindent
$\bullet$ Universal simulation of an arbitrary effective Hamiltonian 
on $\cH_S$ can be implemented in various ways, depending on the set of 
actual simulation requirements and control resources. Suppose that, 
for a given $H_S$, a decoupling set $\cG$ exists, such that $[H_S, 
\Lambda_{\cG}(H_S)]\not =0$, and that periods of free evolution under 
$H_S$ can be alternated with periods of controlled evolution 
under $\Lambda_{\cG}(H_S)$. Then, similarly to the 
twisted decoupler schemes discussed in \cite{prl2}, any Hamiltonian 
$L$ belonging to the Lie algebra generated by $iH_S$, $i\Lambda_{\cG}(H_S)$
can be reached in principle -- implying universal control in the 
generic case \cite{seth}. Even in the unfavorable situation where 
$H_S$ may consist of a single term (say $\sigma_z$ for a qubit), 
arbitrary Hamiltonians can still be engineered if, for instance, both 
a set of b.b. operations averaging $H_S$, and a slow application of 
a Hamiltonian $Z \in \cZ(\cG)$ can be effected. Then one can again 
alternate evolutions under $H_S$ with controlled evolutions where
$Z$ is applied in parallel with the decoupler $\cG$ \cite{prl2}, 
obtaining as above universality in the generic case. (In the qubit 
example, with $H_S=\sigma_z$, one can choose $\cG=\cG_{{\sc cp}_x}$
and apply $Z=\sigma_x$ via weak/slow control \cite{prl2}). Of course, 
such programming procedures may require additional external capabilities 
beyond b.b. control and, in general, they will not ensure universality
starting from an {\sl arbitrary} (possibly unknown) $H_S$. An elegant
approach applicable to this general situation has been recently 
elaborated in \cite{beth}, where the possibility of arbitrary 
universal simulation is related to the identification of 
special decoupling groups called {\sl transformer groups}. 
Starting from any Hamiltonian $A$, decoupling according to a 
transformer group is able to map $A$ into any desired effective 
Hamiltonian. For instance, a transformer group for a two-dimensional
system (a single qubit) is generated under the natural representation 
by the four b.b. operations $\{ i\sigma_x, i\sigma_y,  i\sigma_z, R\}$, 
$R$ being the rotation by $2\pi/3$ about the axis 
$\hat{n}=(1,1,1)/\sqrt{3}$ that cyclically permutes the Pauli matrices.

\section{Encoded dynamical decoupling}

So far, the full $N=2^n$-dimensional state space of $n$ physical qubits 
has been exploited. However, restricting to a $N_L$-dimensional {\sl 
quantum code}, $N_L <N$, may prove extremely useful in QIP -- the 
resulting benefits sometimes largely compensating the overheads and 
complications arising from dealing with a smaller number $n_L<n$ of 
encoded qubits. In particular, the two primary motivations
for seeking appropriate encodings are either to ensure protection 
against noise -- via active error-correcting codes \cite{preskill} 
or passive noiseless codes on decoherence-free subspaces/noiseless
subsystems (DFSs/NSs) \cite{dfstheory,nstheory,qubit,zanardi2} -- or to 
allow for alternative routes to universality based on the physically 
available interactions on appropriately defined subsystem qubits -- in 
the so-called approach of {\sl encoded universality} \cite{qubit,eu}.

A code ${\cal H}_L$ can be generally thought of as a distinguished 
subsystem of the physical state space of $S$, determined by a 
correspondence of the form 
\begin{equation}
\cH_S \simeq \cH_L \otimes \cH_Z \oplus \cR \:,
\label{code}
\end{equation}
for some $\cH_Z, \cR$. $\cH_L$ is the logical (or computational)
factor, which reduces to a proper subspace $\cH_L \subset \cH_S$ 
when the ``syndrome'' co-factor is one-dimensional, $\cH_Z \simeq 
{\bf C}$. The summand $\cR$ collects the non-computational states 
in $\cH_S$. 
A code can be algebraically characterized with respect to a suitable
algebra $\cA$ of operators on $\cH_S$. The prototype example is the
a {\sl noiseless code}, whereby the appropriate algebra
$\cA$ is the associative {\sl interaction algebra} \cite{nstheory}  
containing the complex linear combinations of arbitrary products 
of $H_S$, all (or a subset of, see Sect. V) the error generators 
$S_\alpha$s, and the $\openone$.  
Then in general $\cA$ can be expressed, with respect to an appropriate 
basis in $\cH_S$, as a direct sum of $d_J$-dimensional complex 
matrices, each appearing with a multiplicity $n_J$, 
\begin{equation}
U \cA U^\dagger =  \bigoplus_{J\in {\cJ}} \openone_{n_J} \otimes
\text{Mat}_{d_J}({\bf C})\:,
\label{a}
\end{equation}
where the change of basis $U$ in $\cH_S$ is made explicit, and  
$\sum_{J\in {\cJ}} n_J d_J =N$. With respect to the same basis, 
the algebra $\cA'= \{ X \in \text{End}(\cH_S) \,|\, [X, \cA]=0 \}$
({\sl commutant} of $\cA$ in End$(\cH_S)$ \cite{nstheory}) 
represents as
\begin{equation}
U \cA' U^\dagger = \bigoplus_{J\in {\cJ}} \text{Mat}_{n_J}({\bf C}) 
\otimes \openone_{d_J}\:.
\label{aprime}
\end{equation}
Thus, under the unitary map $U$, the state space $\cH_S$ becomes
isomorphic to   
\begin{equation}
\cH_S \simeq \bigoplus_{J \in {\cJ}} \cC_J \otimes \cD_J 
\simeq \bigoplus_{J\in {\cJ}} {\bf C}^{n_J} \otimes {\bf C}^{d_J} \:,
\label{statespace}
\end{equation}
{\it i.e.} a direct sum of effectively bi-partite subspaces. 
Each of the left (or right) factors in this decomposition can be 
associated with a $\cA$-code (or $\cA'$-code), respectively. Typically,
to obtain a noiseless code (a DFS or a NS), one selects a fixed 
factor $\cC_{J_\ast}=\cH_L$, with dimension $N_L=n_{J_\ast}$ 
-- in which case, with respect to (\ref{code}), $\cH_Z=\cD_{J_\ast}$, 
and $\cR=\oplus_{J \not = J_\ast} \cC_J \otimes \cD_J$. 
Note that, because of (\ref{aprime}), the code is an irreducible 
subspace of the commutant $\cA'$ \cite{nstheory,dygen}.   

If $N_L \geq 2^{n_L}$, then $\cH_L$ can protect the state space of 
$n_L$ logical qubits against noise in $\cA$. While the {\sl global} 
structure of $\cH_L$ is sufficient for establishing storage
or even existential universality results over $\cH_L$, an additional,
crucial requirement on the {\sl local} structure of $\cH_L$ stems from
the tensor product nature of QIP. In other words, an {\sl encoded
tensor product} structure on $\cH_L$ is necessary for addressing 
notions of {\sl efficient} simulation or universality over $\cH_L$. 
This issue is especially evident in the encoded universality 
approach, where the primary algebraic structure one considers is 
the Lie algebra $\cL$ generated under commutation by the set of easily 
implementable Hamiltonians \cite{eu}. (Formally, $\cL$ plays the same
role as $\cA'$ in the noiseless approach.) Even if the code size $N_L$ 
is large enough to accommodate many qubits, without a precise 
mapping that define encoded qubits there is no way for assessing the 
potential of this set of interactions in terms of one- and two-qubit 
encoded gates useful for implementing a quantum circuit.  
In its essence, properly defining this encoded tensor product structure 
is equivalent to properly constructing qubits in a given physical system
\cite{qubit}. 
While no conclusive solution seems available to date, a practically 
motivated approach consists in identifying single-qubits encodings 
into {\sl small} blocks of (2 to 4) physical qubits, and then inducing
a tensor product structure by adjoining (or ``conjoining'') blocks 
\cite{qubit,eu,kempe1}. 
To do so, the system is partitioned into clusters $\{ c_\ell\}$ of physical 
qubits, {\it i.e.} $\cH_S= \prod_\ell \cH^{(c_\ell)}$, and a mapping of the 
form (\ref{code}) defines the state space $\cH_{L_\ell}$ of the $\ell$th 
encoded qubit starting from $\cH^{(c_\ell)}$. An overall structure as 
in (\ref{code}) still emerges with 
\begin{equation}
\cH_L =\cH_{L_1} \otimes \ldots \otimes \cH_{L_{n_L}} \:,
\label{final}
\end{equation}
and $\cR$ grouping all contributions involving $\cR_\ell$ for at least 
one cluster. The code $\cH_L$ can still be algebraically characterized
as being, in general, embedded into one (or more) invariant subspaces
of $\cA'$ -- or $\cL$, as appropriate \cite{eu,kempe1}. 
For definiteness, I focus on the case where encoding is motivated 
by noise protection against $\cA$ henceforth.    
 
Let the set of encoded qubits be specified by encoded qubit 
observables, 
\begin{equation}
\{ \sigma_a^{L_\ell} \}\:,\hspace{1cm} a=x,y,z; \;\; \ell=1,\ldots,n_L\:,
\label{obs0}
\end{equation}
satisfying Pauli-matrices commutation and anti-commutation rules 
\cite{qubit}, and belonging to $\cA'$. The relevant situation for 
introducing {\sl encoded dynamical control} assumes that the 
natural system dynamics 
-- specified by $H_S$ and possibly by some error generators in $\cN$ 
whose effect is not eliminated by the encoding -- are {\sl expressible 
in terms of encoded qubit observables}. The goal is then to actively
turn on/off selected encoded interactions without spoiling the benefits
associated with the underlying encoding. Let $\cU(\cH_L)\simeq 
\cU(2^{n_L})$ denote the group of unitary operators over $\cH_L$.
As in the un-encoded case, the decoupling problem can be defined by a 
discrete set $\cG^L=\{ U_k^L\}$ of b.b. control operations, which 
represent encoded rotations over $\cH_L$, $\cG^L \subset \cU(\cH_L)$.
In the simplest setting, $\cG^L$ will itself form a group of encoded 
rotations ({\sl encoded decoupling group}). 

There are two minimal requirement for an operator $U_k^L$ 
to provide a legitimate unitary transformation over $\cH_L$: \\
\noindent $\bullet$ 
the gate should never draw the system outside the protected region; \\
\noindent
$\bullet$ 
the qubit mapping should be preserved at the end of the gate \cite{note}. 
\\
\noindent
Both conditions are satisfied if $U_k^L$ is generated by a Hamiltonian
$A_k^L$ that is expressible in terms of encoded qubit observables 
-- as it suffices to the present purpose. Accordingly, $U_k^L=
\exp(i \eta \delta A_k^L)$, for effective strength and time parameters
$\eta$, $\delta$, respectively -- in such a way that the limit 
$\eta \rightarrow \infty$, $\delta \rightarrow 0$ 
with a finite b.b. action $\eta\delta$ can be achieved. 
While specifying how rotations are effected is irrelevant in the 
idealized b.b. limit of instantaneous control actions considered so 
far, it clearly becomes important in a realistic scenario where pulses 
have a finite duration and the evolution during the pulses should 
therefore be taken into account. 
Within AHT, compensation schemes have been developed for dealing 
with pulse-length corrections \cite{haeberlen}. For application with 
encoded pulses, it is necessary to preliminarily make sure that the 
benefits of the encoding are not lost during the pulses.
Suppose that unitary operations in $\cU(\cH_S)$ exist, 
whose action is {\sl not} generated by Hamiltonians in $\cA'$, 
but whose {\sl net} effect matches, upon restriction to $\cH_L$, 
the one associated with some $U^L_k$. Then besides the correction 
effects that also appear for $\exp(i \eta \delta A^L_k)$ for finite 
$\delta$ (and $A_k^L \in {\cal A}'$), 
additional errors are associated with the departure 
from $\cH_L$ during $\delta$. While elimination of these effects 
motivates in principle the necessity of using encoded Hamiltonians, 
in practice different compensation techniques may be attempted, for
instance by resorting to robust control design \cite{dfs}. 
 
With these definitions and caveats in mind, the applicability of 
decoupling methods directly carries over to the encoded case, once
qubits and qubit operators are formally replaced with their encoded 
counterparts. 
Thus, if an operator $X=F[\{\sigma_a^j\}]$ has a given 
structure in terms of physical qubit observables, and a decoupling 
scheme according to $\cG=\{U_k\}$ accomplishes a desired averaging 
effect via
\begin{equation}
\Lambda_{\cG}(X)=\sum_{k \in {\cal K}} \tau_k U_k^\dagger X U_k \:, 
\end{equation}  
then an equivalent averaging effect is obtained on an encoded operator
$X^L$ with the same functional dependence $X^L=F[\{\sigma_a^{L_\ell}\}]$ 
on encoded qubit observables, via the encoded quantum operation
\begin{equation}
\Lambda_{\cG^L}(X^L)=\sum_{k \in {\cal K}} \tau_k U_k^{L\,\dagger} X^L 
U_k^{L} \:. 
\end{equation}

\section{Engineering of encoded closed-system dynamics} 

Suppose that the noise generated by operators in $\cN$ is fully 
taken care of by the chosen encoding, or that noise is not a concern 
to begin with, as in the encoded universality approach. Then encoded
dynamical decoupling may provide a tool for encoded universal 
Hamiltonian simulation. 

If, as assumed above, $H_S$ expresses in terms of encoded observables, 
then the natural dynamics already implements a non-trivial logical 
transformation over the code. This provides the primary input to be 
exploited for engineering desired effective evolutions by encoded 
manipulations. As in the un-encoded case, detailed knowledge on the 
structure of $H_S$ may or may not be a data of the problem. For 
{\sl arbitrary} $H_S$, the results established in \cite{beth} imply 
that universal encoded simulation is achievable, in principle, if a 
finite encoded transformer can be constructed. The transformer groups 
so far identified \cite{beth} could be useful in principle for code 
size $N_L=2,3$ -- although, unfortunately, practical impact is
limited by the large number of encoded rotations involved 
($|\cG^L|=24$ for the above-mentioned single-qubit transformer, 
and $|\cG^L| \geq 168$ for dimension 3). 

For the less ambitious target of generating a universal set of 
encoded evolutions starting from a given (known) $H_S$, simpler
encoded programming strategies along the lines sketched in Sect. II
may suffice. For instance, in the generic case, the Hamiltonian $H_S$,
together with a non-commuting Hamiltonian $\Lambda_{\cG^L}(H_S)$
obtained from $H_S$ via some encoded decoupling procedure, will 
fulfill the conditions for generating the whole Lie algebra 
u($\cH_L)$ of anti-Hermitian encoded Hamiltonians  -- thereby
implying universality over $\cH_L$, at least at the existential 
level. The required group $\cG^L$ of encoded b.b. rotations may
be as simple as an encoded Carr-Purcell-type group. Let, for 
instance, $\pi^L$ denote an encoded $\pi$ rotation (acting on 
one or more qubits), and $\cG^L_{\sc cp}$ the associated encoded 
group. Then it is always possible to separate terms in $H_S$ which 
are symmetric $(s)$ and anti-symmetric $(a)$ under $\cG_{\sc cp}^L$,
\begin{equation}
H_S= H_S^s + H_S^a\:, 
\end{equation}
with 
\begin{equation}
\pi^L \,H_S^{(s)} \,\pi^L = H_S^s = \Pi_{\cG^L}(H_S)\:, \hspace{1cm}
\pi^L \,H_S^{(a)} \,\pi^L = - H^a_S \:.
\end{equation}
Thus, the above argument generally applies provided 
$[H_S, \Pi_{\cG^L}(H_S)]=[H_S^s, H_S^a] \not =0$, and the two 
Hamiltonians $H_S$, $\Pi_{\cG_{\sc cp}}(H_S)$ accordingly suffice for 
universal encoded control. An experimental demonstration using a 
DFS-qubit encoded into the zero-quantum subspace of two nuclear spins 
is reported in \cite{dfs}.

\subsection{Example: A single NS-encoded qubit}

A similar procedure could be relevant for obtaining universal 
encoded control over a NS-encoded qubit of three spin 1/2 under 
general collective noise \cite{nstheory}. 
Suppose that the noise generators $S_\alpha$ are the global, 
permutation-invariant Pauli operators $S_a =\sum_{j=1}^3
\sigma^j_a$, $a=x,y,z$, and that the natural system Hamiltonian has
the isotropically-coupled form 
\begin{equation}
H_S= \Omega S_z + J_{12} s_{12}  + J_{23} s_{23}  + J_{31} s_{31}\:, 
\label{nsham}
\end{equation}
where the Heisenberg exchange coupling $s_{jk} = \vec{\sigma}^j 
\cdot \vec{\sigma}^k$, and the parameters $\Omega, J_{jk}$ are real.
Then $\cA$ is the algebra of completely symmetric operators over 
$\cH_S \simeq ({\bf C}^2)^{\otimes 3}$, and $\cA'$ can be identified 
with the group algebra of the permutation group $\cS_3$ (under the 
natural representation in $\cH_S$). A NS code under $\cA$ is 
identified by a correspondence of the form (\ref{code}) -- 
with $\cH_Z \simeq {\bf C}^2$ carrying the irreducible representation 
of su(2) corresponding to total angular momentum $J=1/2$, 
$\cH_L \simeq {\bf C}^2$ carrying the two-dimensional irreducible 
representation of $\cS_3$, and $\cR \simeq {\bf C}^4$ being the 
invariant subspace of states with total angular momentum $J=3/2$ 
\cite{nstheory,qubit,kempe}.
Explicit expressions for encoded qubit observables (and the associated
logical states) are given in \cite{qubit}. In particular,
\begin{equation}
\sigma^L_x =_L {1 \over 2} \Big( \openone + s_{12} \Big)\:,
\hspace{1cm} 
\sigma^L_y =_L -{\sqrt{3} \over 6} \Big( s_{23} - s_{31} \Big)\:.
\end{equation}
This allows to rewrite the Hamiltonian (\ref{nsham}), up to 
irrelevant contributions which are constant over $\cH_L$, as 
\begin{equation}
H_S =_L (2J_{12} - J_{23} -J_{31}) \, \sigma^L_x 
+ \sqrt{3} \,(J_{31} -J_{23})\, \sigma^L_y \:.
\label{ns2}
\end{equation}
Note that the vanishing of $H_S$ for a fully symmetric coupling
network, $J_{jk}=J$ $\forall j,k$, correctly verifies the identity 
action of permutation-invariant operators over $\cH_L$ ($H_S \in 
\cA$ in this case). Given $H_S$ in the form (\ref{ns2}), the 
above-mentioned universality scheme based on 
encoded Carr-Purcell sequences becomes applicable in principle 
provided one has to ability to enact rapid, encoded $\pi^L$ pulses.
Because, with respect to the chosen code, the action of the 
$\sigma_x^L$ operator is effectively identical to swapping the 
physical qubits 1 and 2, this is for instance achievable if the 
exchange Hamiltonian $s_{12}$ can be switched on for the appropriate 
time. As a result, one generates an effective encoded Hamiltonian 
$\Pi_{\cG^L_{ {\sc cp}_x} }(H_S)= (2J_{12} - J_{23} -J_{31} ) 
\sigma^L_x$ which, together with $H_S$, allows for universality.
Although not directly applicable to the weakly-coupled molecule 
used in \cite{ns} to realize the above NS, and generally demanding
for NMR implementations as the $J$-coupling parameters are not 
controllable, these ideas could prove viable for solid-state
implementations where Heisenberg interactions are in principle 
fully tunable \cite{eu}. Explicit universality constructions 
relevant to quantum computing architectures based on 
exchange interactions are provided in \cite{wu}.

\subsection{Example: A pair of DFS-encoded qubits}

As a further example, motivated from NMR, imagine two pairs of
spin 1/2 nuclei, corresponding to different species (Hydrogen and 
Carbon, for instance), subjected to block-collective dephasing noise.
Let the two clusters be associated with the pairs $(1,2)$ and $(3,4)$, 
respectively. Then the two error generators are $S_z^{(c_1)}= 
\sigma_z^1+ \sigma_z^2$, $S_z^{(c_2)}= \sigma_z^3+ \sigma_z^4$, and
$\cA= \cA_z^{(c_1)}\otimes \cA_z^{(c_2)}$ in terms of the interaction 
algebras $\cA_z^{(c_\ell)}$ generated by $S_z^{(c_\ell)}$ for dephasing 
on 2 qubits \cite{dfs}. 
Protection against errors in $\cA$ can be accomplished by replacing each 
physical pair with a DFS-encoded qubit supported by the states having 
$S_z^{(c_1)}=0$ and $S_z^{(c_2)}=0$ {\it i.e.},
\begin{equation}
|i_{L_1} j_{L_2}\rangle= |i_{L_1}\rangle\otimes |j_{L_2}\rangle \:,
\hspace{1cm} |0_{L_\ell}\rangle= |01\rangle^{(c_\ell)}\:, \;
|1_{L_\ell}\rangle= |10\rangle^{(c_\ell)}\,, \; \ell=1,2\:.
\label{enc}
\end{equation} 
In this case, with reference to the general structure (\ref{code}), 
each DFS encoding has the form $\cH^{(c_\ell)} \simeq \cH_{L_\ell}
\oplus \cR_\ell$, with $\cH_{L_\ell}=\text{span}\{|0_{L_\ell}\rangle,
|1_{L_\ell}\rangle \} \simeq {\bf C}^2$, $\cR_{\ell}=\text{span}
\{|00\rangle_\ell, |11\rangle_\ell \} \simeq {\bf C}^2$ 
($\cH_{Z_\ell}\simeq {\bf C}$ is irrelevant for subspace-encodings). 
Correspondingly, for 4 spins
$\cH_L\simeq {\bf C}^2 \otimes {\bf C}^2$ is a 4-dimensional subspace 
of the 6-dimensional zero-quantum subspace corresponding to total 
$z$ angular momentum $S_z=\sum_j \sigma^j_z=0$, $j=1,\ldots,4$, and $\cR$ 
collects all contributions where $S_z^{(c_\ell)} \not =0$ for at least 
one pair. Encoded observables for the above qubits are provided, for
example, by the choice \cite{dfs}
\begin{equation}      
\sigma_z^{L_1} =_L {1 \over 2} \Big( \sigma_z^{1} -\sigma_z^{2} \Big)\:,
\hspace{1cm}
\sigma_x^{L_1} =_L {1 \over 2} \Big( \sigma_x^{1}\sigma_x^{2} + 
\sigma_y^{1}\sigma_y^{2} \Big)  \:,
\label{observ}
\end{equation}
and similarly for qubit $L_2$. 

In general, the physical spin system will exhibit a strongly-coupled 
spectrum described by an internal Hamiltonian composed of both Zeeman 
spin-field and indirect spin-spin interactions, 
\begin{equation}
H_S= \sum_{j=1,\ldots,4} \pi \nu_j \sigma_z^j +
\sum_{j<j'=1,\ldots,4} {\pi \over 2} J_{jj'} \vec{\sigma}^j \cdot 
\vec{\sigma}^{j'} = \sum_j H_{Z_j} + \sum_{j<j'} H_{J_{jj'}}\:,
\label{hnmr}
\end{equation}
where the chemical shifts and $J$-coupling parameters are understood 
in frequency units. For usual values of the static Zeeman field, the 
contribution to the total energy of a given pair of spins $(j,j')$
due to the coupling $H_{J_{jj'}}$ can be treated as a perturbation
with respect to $H_{Z_{j (j')}}$ -- the diagonal $\sigma_z^j 
\sigma_z^{j'}$ and off-diagonal 
$\sigma_x^j\sigma_x^{j'}+\sigma_y^j\sigma_y^{j'}$ 
terms leading to first- and second-order correction effects in 
$J_{jj'}$, respectively. Because the differences in the chemical 
shifts $|\nu_j-\nu_{j'}|$ are larger when different nuclear species 
are involved, the approximation of neglecting off-diagonal 
couplings (weak-coupling limit) is well justified for hetero-nuclear 
interactions. Thus, (\ref{hnmr}) can be effectively replaced by
\begin{equation}
H_S= \sum_{j=1,\ldots,4} \pi \nu_j \sigma_z^j + {\pi \over 2} 
\left( J_{12} \vec{\sigma}^1 \cdot 
\vec{\sigma}^{2}  + J_{34} \vec{\sigma}^3 \cdot \vec{\sigma}^{4} \right) + 
\sum_{j=1,2;j'=3,4} {\pi \over 2} J_{jj'} \sigma_z^j \sigma_z^j \:.
\label{hnmr2}
\end{equation}
In terms of the encoded qubit observables given in (\ref{observ}), 
the chemical shift terms immediately rewrite as combinations of 
$S_z^{(c_\ell)}$ operators (constant over $\cH_L$) and logical 
$\sigma_z^{L_\ell}$ operators, whereas the homo-nuclear spin-spin
interactions contribute with logical $\sigma_x^{L_\ell}$ operators
(and additional constant terms proportional to $\sigma_z^j 
\sigma_z^j$). The remaining hetero-nuclear bilinear couplings 
in (\ref{hnmr2}) give
\begin{equation}
A\, S_z^{(c_1)}S_z^{(c_2)} + B\,  S_z^{(c_1)}\sigma_z^{L_2} + 
C\, S_z^{(c_2)}\sigma_z^{L_1} + D\, \sigma_z^{L_1}\sigma_z^{L_2}\:,
\end{equation}
with coefficients 
\begin{eqnarray}
A &=& {1 \over 8} \Big( J_{13}+  J_{14}+ J_{23}+ J_{24} \Big) \:,
\hspace{1cm}
B = {1 \over 4}\Big ( J_{13}- J_{14}+ J_{23}- J_{24} \Big) \:,
\nonumber \\
C &=& {1 \over 4}\Big ( J_{13}+  J_{14}- J_{23}- J_{24}\Big) \:,
\hspace{1cm}
D = {1 \over 4} \Big( J_{13}- J_{14}- J_{23}+ J_{24} \Big) \:.
\end{eqnarray}
Again, an overall identity action over $\cH_L$ is obtained for 
a symmetric coupling network -- as well as for other special 
coupling patterns in the system.  Assuming that none of these 
non-generic circumstances is met, the action of the overall 
Hamiltonian $H_S$ over the code is finally:
\begin{equation}
H_S=_L \pi\, \Big( \Delta \nu_{12} \,\sigma_z^{L_1} +  \Delta \nu_{34} 
\, \sigma_z^{L_2} + J_{12} \,\sigma_x^{L_1} + J_{34} \,\sigma_x^{L_2}
+ D  \,\sigma_z^{L_1}\sigma_z^{L_2}\Big) \:, 
\label{finalh}
\end{equation}
with $\Delta \nu_{12}= \nu_1-\nu_2, \,\Delta \nu_{34}= \nu_3-\nu_4$, 
respectively.
 
Remarkably, the structure of the Hamiltonian (\ref{finalh}) for the 
two encoded qubits is very similar to the one of the Hamiltonian 
describing two weakly-coupled physical spins -- except, in the 
ordinary NMR setting, control along the transverse $\sigma_x$ (or 
$\sigma_y$) directions is directly supplied by external radio-frequency 
fields. While the potential of (\ref{finalh}) for universal encoded control 
should be expected by analogy with the un-encoded case, various schemes
may be specified depending on the actual control capabilities.  
At the existential level, universality follows from the ability of 
rapidly effecting a single encoded rotation, say a ``hard'' 
(non-selective) encoded $\pi^L$ pulse, 
$\pi^L_x=\pi^{L_1}_x\pi^{L_2}_x$  about the encoded $x$ axis. 
Because the associated averaging produces the encoded 
effective Hamiltonian
$\Pi_{\cG^L}(H_S)= J_{12}\sigma_x^{L_1} + J_{34}\sigma_x^{L_2}+ D 
\sigma_z^{L_1}\sigma_z^{L_2}$, by appropriately alternating evolution 
periods under $H_S$ and $\Pi_{\cG^L}(H_S)$, arbitrary encoded 
Hamiltonians can be enacted in principle 
through repeated commutation \cite{seth}. 

Constructive prescriptions become possible as soon as a wider range
of control options is accessible. In particular, schemes for effecting 
a universal set of encoded gates -- including $(i)$ all single qubit 
rotations and $(ii)$ a two-qubit phase coupling -- can be provided. 
Assume that, in formal analogy with the standard NMR setting, the 
parameters $J_{12}$ and $J_{34}$ are independently tunable at will. 
To selectively rotate each encoded qubit about the $x$ axis, turn on 
the appropriate $\sigma^{L_\ell}_x$ term in (\ref{finalh}) and
simultaneously refocus $\sigma^{L_\ell}_z$ evolutions by applying 
encoded $\pi_x^{L_1}\pi_x^{L_2}$ pulses as above. 
Note that although this action preserves the 
$\sigma^{L_1}_z \sigma^{L_2}_z$ coupling, the corresponding 
evolution can be neglected under the usual assumption that single-qubit 
rotations can be effected rapidly enough. In fact, the whole evolution 
induced by (\ref{finalh}) can be neglected during the $x$ rotation if 
the strength of the appropriate  $J_{12}$ or $J_{34}$ parameter can be 
arbitrarily controlled as assumed. 
Also, note that the same sequence of hard $\pi_x^{L_1}\pi_x^{L_2}$ pulses 
allows to selectively leave this phase coupling alone if both $J_{12}$ 
and $J_{34}$ are kept to zero -- except during the (infinitesimal) 
durations of the refocusing pulses.   
Selective rotations about the appropriate encoded $z$ axis are only 
slightly more demanding, as they require the hard logical $\pi^L_x$ 
pulses to be replaced by ``soft'' encoded $\pi^{L_\ell}_x$ on the 
intended spin alone.
The relevance of these and related universality constructions has 
been highlighted in \cite{wu}.

However, the above tunability requirements are too stringent,
for instance, for NMR implementations -- where the natural Hamiltonian
(\ref{finalh}) is {\sl always} active, and interactions can be only 
effectively set to zero via appropriate refocusing. 
Additional constraints may arise for typical implementation 
parameters. In particular, one may expect that the encoded chemical 
shift evolutions will dominate in (\ref{finalh}), {\it i.e.}  
$\Delta \nu_{12}, \Delta \nu_{34} \gg J_{12}, J_{34} \gg D$. 
Nevertheless, universal control can still be gained by having access to 
a small set of encoded operations. No ability of arbitrarily tuning 
encoded Hamiltonians is assumed, apart from the {\sl fixed} combinations 
of strength and times corresponding to logical $\pi^L$ pulses. However, 
unlike the previous case, encoded $\pi^L$ rotations about {\sl two} axes 
are generally needed. The encoded decoupling schemes for implementing 
the required universal set of coupling are as follows. Selective 
rotations about the appropriate $z$ axis are easiest, as one can take 
advantage of the natural averaging of the $\sigma^{L_\ell}_x$ and 
$\sigma^{L_1}_z\sigma^{L_2}_x$ that is enforced by the above hierarchy. 
One need only refocus the undesired encoded phase evolution (say, 
$\sigma^{L_2}_z$)
by applying sequences of encoded selective $\pi^{L_\ell}_x$ pulses 
(say, $\pi^{L_2}_x$).
It is easily seen that decoupling purely based on simultaneous 
$\pi^{L_1}_x\pi^{L_2}_x$ pulses as 
above is not useful, without the ability to separately controlling
the coefficients in the resulting effective Hamiltonian, to implement 
selective $x$ rotations or the two-qubit coupling. 
However, a selective implementation of, say, 
$\sigma_x^{L_1}$ can be engineered, for instance, by the encoded 
pulse sequence
\begin{equation}
\pi^{L_1}_x\pi^{L_2}_x \,- \Delta t - \, \pi^{L_1}_x\pi^{L_2}_z \,-
\Delta t - \,  
\pi^{L_1}_x\pi^{L_2}_x \,- \Delta t - \, \pi^{L_1}_x\pi^{L_2}_z \:, 
\hspace{1cm}\Delta t= T_c/4\:,
\label{s1}
\end{equation}
which corresponds to subjecting qubit 2 to maximal encoded averaging 
according to 
$\cG^L_{max}=\{ \openone^L, \sigma^L_x,  \sigma^L_y,  \sigma^L_z \}$, 
and cycling twice qubit 1 through the encoded Carr-Purcell sequence 
refocusing $\sigma_z^{L_1}$.  A similar sequence works for 
rotating the second encoded qubit, by interchanging 1 and 2 in 
(\ref{s1}). If all single-qubit interactions are refocused instead, 
by applying the above $\cG^L_{max}$ to both qubits, 
\begin{equation}
\pi^{L_1}_x\pi^{L_2}_x \,- \Delta t - \, \pi^{L_1}_z\pi^{L_2}_z \,-
\Delta t - \,  
\pi^{L_1}_x\pi^{L_2}_x \,- \Delta t - \, \pi^{L_1}_z\pi^{L_2}_z \:, 
\hspace{1cm}\Delta t= T_c/4\:,
\end{equation}
the $\sigma^{L_1}_z \sigma^{L_2}_z$ evolution is selectively 
extracted from (\ref{finalh}). 

The problem of how to actually effect the required 
logical $\pi^L$ rotations remains challenging in reality, as the 
evolutions induced by the available radio-frequency fields do not
in general correspond to encoded Hamiltonians. Thus, even though
it is always possible to enforce a unitary propagator whose net 
action is the same as the one of the required encoded rotation, 
the corresponding sequence of physical gates may be complicated 
and, as noted above, will cause departure from the code during the 
finite pulsing time. 
A possible way out is explored in \cite{dfs}, based on the idea
of compensating for the resulting exposure to noise by optimizing 
the length of the relevant control sequences and by incorporating 
intrinsic robustness features using composite pulse techniques 
\cite{composite}. The situation is relatively straightforward for 
the special case of $\pi^L$ pulses, as the encoded rotations they
effect have a simple translation in terms of realizable physical
spin propagators. The simplest instance is the action of a 
$\pi^{L_1}_x\pi^{L_2}_x$, which is identical to the action induced 
over $\cH_L$ by a hard $\pi_x$ pulse on all four spins. In fact, 
the following correspondence for the logical $\pi^L$ pulses 
involved in the above encoded sequences holds (up to irrelevant 
phase factors):
\begin{eqnarray*}
\pi^{L_1}_x \;\;& \leftrightarrow &  \;\;\sigma_x^1\sigma_x^2 \:, \\
\pi^{L_2}_x  \;\;& \leftrightarrow & \;\; \sigma_x^3\sigma_x^4 \:, \\
\pi^{L_1}_x \pi^{L_2}_x  \;\;& \leftrightarrow & \;\; \sigma^1_x\sigma^2_x
\sigma_x^3\sigma_x^4 \:, \\ 
\pi^{L_1}_x \pi^{L_2}_z \;\;  & \leftrightarrow &\;\; \sigma^1_x\sigma^2_x
\sigma_z^4 \:, \\ 
\pi^{L_1}_z \pi^{L_2}_x \;\;  & \leftrightarrow &\;\; \sigma^2_z\sigma^3_x
\sigma_x^4 \:, \\ 
\pi^{L_1}_z \pi^{L_2}_z \;\;& \leftrightarrow & \;\;\sigma^2_z\sigma_z^4 \:. 
\end{eqnarray*} 
Accordingly, universal manipulation of encoded evolutions with reduced 
error rate is still achievable if robust ways for effecting the above 
physical operations can be devised \cite{dfs,fastpulse}.

\section{Engineering of encoded open-system dynamics} 

A further direction for application is suggested by looking at 
encoded dynamical decoupling as a tool for encoded quantum error 
suppression. Similar to the idea of concatenating passive noise
protection via DFSs/NSs with quantum error correction -- by 
operating error-correcting codes directly on DFS/NS-encoded qubits
\cite{concat}, the idea is now to concatenate passive noise 
protection with active dynamical control -- by effecting error 
suppression directly on encoded qubits.

\subsection{Example: A decohering encoded qubit}

By analogy to the prototype example of a single decohering 
qubit analyzed in \cite{pra}, the prototype situation is provided 
by a single decohering {\sl encoded} qubit. 
Imagine a system $S$ composed of two physical qubits, which are 
subjected to a purely dephasing interaction as a result of the 
coupling to two bosonic environments $B_1$, $B_2$ {\it i.e.}, the 
overall system is described by a Hamiltonian of the form
\begin{equation}
H = H_S+H_{B_1}+H_{B_2}+H_{SB_1} +H_{SB_2} \:, 
\label{htot}
\end{equation}
for uncoupled subsystem's Hamiltonians
\begin{equation}
H_S= \sum_{j=1,2} {\omega_k \over 2} \sigma_z^{j}\:, \hspace{1cm}
H_{B_1}+H_{B_2} = \sum_{k} \omega_k b^{(1)}_k b^{(1)\dagger}_k +
\sum_{\ell} \omega_\ell b^{(2)}_\ell b^{(2)\dagger}_\ell \:,
\end{equation}
and a total interaction Hamiltonian
\begin{equation}
H_{SB_1} +H_{SB_2}= \sum_{j=1,2} \sigma_z^{j} \otimes
\Big\{ \sum_k g_{k j}^{(1)}(b_k^{(1)} + b_k^{(1)\dagger}) +
       \sum_\ell g_{\ell j}^{(2)}(b_k^{(2)} + b_k^{(2)\dagger})  
\Big\} \:.
\label{hint}
\end{equation}
In the above equations, $\omega_j$, $j=1,2$, are the natural frequencies 
of single-qubit evolutions, $b_k^{(1)}, b_k^{(1)\dagger}$ are bosonic 
operators for the environment $B_1$ (similarly for $B_2$), and the 
parameters $g_{k j}^{(1)}$ ($g_{\ell j}^{(2)}$) determine the coupling 
strength of qubit $j$ to mode $k$ of bath 1 (or mode $\ell$ of bath 2, 
respectively). Identity operators on the appropriate subsystems are
also understood as needed.  
Then implementing decoupling according to the group ${\cG}_{{\sc cp}_x}=
\{\openone, \sigma_x^{1} \sigma_x^{2} \}$ would 
dynamically suppress both error generators -- thereby allowing to 
reduce the dephasing noise on both qubits to a level, in principle,
as low as desired. However, as recalled earlier, this is only effective 
in practice in the limit of fast control where \cite{pra,tombesi}
\begin{equation}
T_c \lesssim \text{min}_i \{ \tau_c^{(i)} \} \:, 
\end{equation}
$\tau_c^{(i)}$ denoting the (shortest) correlation time of the $i$th 
environment. Suppose now that $B_2$ is sufficiently ``slow'', but $B_1$
is ``fast'', so that decoupling at the rate determined by
$\tau_c^{(1)}$ becomes unfeasible. Although implementing decoupling
at the slower rate set by $B_2$ reduces the error rate originating from
$B_2$, there is no actual guarantee that the overall error rate is
suppressed due to the possibility of decoherence acceleration from 
modes at frequency higher than $\sim 1/\tau_c^{(2)}$ 
in $B_1$ \cite{pra}.
In any event, both qubits would remain effectively exposed to noise. 
Are there other noise control options worth being considered?

For arbitrary dephasing and just two qubits, quantum error correction 
does not help, neither does passive noise protection -- unless some
symmetries can be identified in the noise. Suppose that, in the 
above interaction Hamiltonian (\ref{hint}), 
$g_{k 1}^{(1)}= g_{k 2}^{(1)}=g_k^{(1)}$ to a good accuracy. 
Then the overall coupling rewrites as
\begin{equation}
H_{SB_1} +H_{SB_2}= S_z \otimes {\cal B}_z^{(1)} + 
\sigma_z^{1} \otimes  {\cal B}_1^{(2)} + 
\sigma_z^{2} \otimes  {\cal B}_2^{(2)} \:,
\label{hint2}
\end{equation}
for $S_z=\sigma_z^{1}+\sigma_z^{2}$ as earlier defined, and appropriate 
environment operators 
${\cal B}_z^{(1)}= \sum_k g_{k}^{(1)}(b_k^{(1)} + b_k^{(1)\dagger})$, 
${\cal B}_j^{(2)}= \sum_\ell g_{\ell j}^{(2)}(b_\ell^{(2)} + 
b_\ell^{(2)\dagger})$, $j=1,2$. Physically, this situation corresponds 
to a noise process consisting of both ``fast'' collective dephasing, due 
to $B_1$, and ``slow'' independent dephasing, due to $B_2$. Whenever a 
symmetry in the open-system dynamics is present, major gains should be
expected by seeking for an appropriate encoding into a DFS or a NS.
In our case, the relevant DFS is the one already introduced in Sect. IVB
{\it i.e.}, the one spanned by states corresponding to total zero 
angular momentum along $z$, $S_z=0$. Thus, a logical DFS qubit is 
defined by the encoding given in (\ref{enc}) (for just one qubit), 
and 
\begin{equation}
{\cal H}_L = \text{span} \{|0_L\rangle, |1_L\rangle \} =
\text{span} \{|01\rangle, |10\rangle \} \:. \label{qubit}
\end{equation}
By construction, this 
encoded qubit is perfectly protected (with infinite distance) against
the noise due to $B_1$. However, mixing with the degrees of freedom 
of $B_2$ is still induced through the error generators 
$\sigma_z^{j}$, $j=1,2$ in (\ref{hint2}). The key observation is 
that this residual noise preserves the coding space, corresponding
to a purely decohering coupling between the encoded qubit and $B_2$. 
Because, by using (\ref{observ}), the error generators simply express 
in terms of encoded observables
\begin{equation}
\sigma_z^{1} =_L  {S_z \over 2} + \sigma_z^L \:, \hspace{1cm}
\sigma_z^{2} =_L  {S_z \over 2} - \sigma_z^L \:,
\label{obs}
\end{equation}  
the action of the total Hamiltonian $H$ (\ref{htot}) on ${\cal H}_L$ 
finally rewrites as 
\begin{equation}
H =_L {\Delta \omega} \,\sigma_z^L + H_B +  
\sigma_z^L \otimes {\cal B}_z \:,
\label{hintfin}
\end{equation}
where $ \Delta \omega = (\omega_1-\omega_2)/2$, $H_B= H_{B_1}+H_{B_2}$, 
and ${\cal B}_z = ({\cal B}_1^{(2)} - {\cal B}_2^{(2)})/2$. This form 
makes it clear that the action of $H$ on the encoded qubit is 
formally identical to the action of the diagonal spin-boson Hamiltonian 
on the physical qubit considered in \cite{pra,leggett}. 

By the same argument valid for suppressing decoherence on a single
physical qubit, implementing a sequence of equally spaced, encoded 
$\pi_x^L$ pulses sufficiently fast with respect to the {\sl slower} 
rate determined by $1/\tau_c^{(2)}$ is now guaranteed to suppress the 
encoded error rate by a factor of (at least) $O[(T_c/\tau_c^{(2)})^2]$
\cite{prl1}.  This corresponds to encoded dynamical decoupling 
according to ${\cal G}^L_{{\sc cp}_x}= \{ \openone^L, \sigma_x^L \}$,  
where $\sigma^L_x$ is again given in (\ref{observ}).
Thus, if the required encoded $\pi^L_x$ rotation can be effected, 
suppression of the ``slow'' dephasing from $B_2$ can be accomplished 
without re-introducing exposure of the encoded qubit to the  
``fast'' dephasing from $B_1$.

\subsection{Generalizations}

Some  generalizations of the above example are worth pointing out.
First, the same encoded decoupling scheme is effective at suppressing
encoded phase errors in a situation where the internal two-qubit 
Hamiltonian is governed by a generally anisotropic exchange Hamiltonian 
\begin{equation}
H_S={\omega_1 \over 2} \sigma_z^{1} + {\omega_1 \over 2} \sigma_z^{2}
+ J_{12}^x \sigma_x^{1}\sigma_x^{2} + 
J_{12}^y \sigma_y^{1}\sigma_y^{2} +
J_{12}^z \sigma_z^{1}\sigma_z^{2} \:,
\end{equation}
with $J_{12}^x= J_{12}^y=J$ and either $ J_{12}^z$ arbitrary, 
$J_{12}^z=0$, or $J_{12}^z=J$ (the so-called $XXZ$, $XY$, and isotropic 
models, respectively \cite{wu}). In the isotropic case in particular, 
which was also examined in Sect. IV and is directly relevant to NMR QIP, 
the encoded open-system Hamiltonian (\ref{hintfin}) is modified to an 
encoded spin-boson Hamiltonian \cite{leggett} as
\begin{equation}
H'=_L {\Delta \omega } \,\sigma_z^L + J\sigma^L_x + H_B +  
\sigma_z^L \otimes {\cal B}_z \:.
\label{hintfin2}
\end{equation}
Once encoded suppression of the $\sigma^L_z$ error generator is 
accomplished as above, control according to the $\sigma_z^L$ Hamiltonian
can be re-introduced, if desired, by implementing one of the 
programming schemes described in \cite{prl2} -- implying the possibility 
of retaining universal noise-suppressed encoded control. 

Situations involving hybrid
dephasing processes with highly correlated components and slow residual 
noise in multi-qubit systems may be also relevant to NMR. If, for instance,
in a four-qubit system, dephasing from $B_1$ is pair-wise correlated on 
qubits (1,2) and (3,4) as considered in Sect. IVB, then 
encoding into the tensor product of the two DFSs described there 
ensures protection against $B_1$. Once this is done, 
applying sequences of encoded $\pi_x^{L_1}\pi_x^{L_2}$
pulses at the appropriate rate causes a suppression of any 
(independent or correlated) residual dephasing due to $B_2$. 

Finally, similar ideas may be more generally applicable to situations 
where the encoded error generators for the residual noise are expressible 
through encoded observables. If, for instance, 
starting from the above two-qubit Hamiltonian (\ref{htot}), the 
action over $\cH_L$ may be cast into the form 
\begin{equation}
H =_L {\Delta \omega } \,\sigma_z^L + H_B +  
\sigma_z^L \otimes {\cal B}_z + \sigma_x^L \otimes {\cal B}_x\:,
\label{multi}
\end{equation}
for appropriate operators $ {\cal B}_{z,x}$ on environment $B_2$, then 
the encoded qubit suffers from depolarizing noise. The resulting error
rate can be suppressed, in principle, by using the encoded annihilator
${\cal G}^L_{max}= \{ \openone^L,\sigma_x^L, \sigma_y^L, \sigma_z^L \}$
introduced above.  
Note that, in terms of coupling to the physical degrees of freedom, 
the interaction (\ref{multi}) involves {\sl multiple-qubit} excitations 
as error generators. Interestingly, system-bath couplings allowing for
similar multi-qubit processes to lowest order are the only class of 
interactions, in addition to the ones exhibiting spatial symmetry, for 
which DFSs are known to exist \cite{bacon}.

To summarize, if noise has both slow and fast components, so that 
achieving full decoupling of the physical degrees of freedom becomes
unfeasible, then encoded dynamical decoupling may be useful in 
situations where a dominant symmetry in the fast noise can be 
exploited to obtain encoded qubits, and the generators for the 
residual slow noise are expressible in terms of encoded observables. 
In this event, using encoded rather than physical
degrees of freedom allows the system to benefit already from enacting 
decoupling operations at the slower rate. However, this looser constraint
on control time scales competes, as already emphasized, with tighter 
symmetry constraints on the useful control Hamiltonians --  as encoded 
Hamiltonians are demanded, and they are not always easily available in 
physical systems.

\section{Conclusions}

I have analyzed the relevance and potential usefulness of active 
dynamical control, as inspired by NMR spectroscopy, in 
the light of the QIP-motivated notion of encoded degrees of freedom. 
Ultimately, the resulting approach of 
concatenating active control with encoded qubits naturally stems
from the program of regarding the information-carrying subsystems
as the primary degrees of freedom for realizing QIP. 
Applications of encoded dynamical decoupling to control both 
Hamiltonian and non-Hamiltonian encoded evolutions have been 
considered in principle. 
While the actual viability within both settings strongly depends on 
the details of the specific implementation and the available control 
resources, I hope that these ideas will motivate further investigation 
and serve as guiding principles to further expand our capabilities to 
manipulating quantum systems and quantum information.

\section{Acknowledgments}

This work was supported from the DOE, under contract W-7405-ENG-36, 
and from the NSA. I am indebted to David Cory for having introduced me
to the richness and beauty of the NMR field. It is a pleasure to thank 
Evan Fortunato for invaluable discussions on these subjects and a 
critical reading of the manuscript.



\begin{thebibliography}{99}


\bibitem{haeberlen0} U. Haeberlen and J. S. Waugh, Phys. rev. {\bf 175}, 
453 (1968). 

\bibitem{haeberlen} U. Haeberlen, {\it High Resolution NMR in Solids: 
Selective Averaging} (Academic Press, New York, 1976).

\bibitem{ernst} R. R. Ernst, G. Bodenhausen, and A. Wokaun, {\it Principles
of Nuclear Magnetic Resonance in One and Two Dimensions} (Oxford University
Press, Oxford, 1994).

\bibitem{cory} D. G. Cory, M. D. Price, and T. F. Havel, Physica D
{\bf 120}, 82 (1998); 
D. G. Cory {\it et al.}, Fortschr. Phys. {\bf 48}, 
875 (2000), and references therein.

\bibitem{pra} L. Viola and S. Lloyd, Phys. Rev. A {\bf 58}, 2733 
(1998).

\bibitem{prl1} L. Viola, E. Knill, and S. Lloyd, Phys. Rev. Lett. 
{\bf 82}, 2417 (1999). 

\bibitem{zanardi1} P. Zanardi, Phys. Lett. A {\bf 258}, 77 (1999).

\bibitem{prl2} L. Viola, S. Lloyd, and E. Knill, Phys. Rev. Lett.
{\bf 83}, 4888 (1999). 

\bibitem{dygen} L. Viola, E. Knill, and S. Lloyd, Phys. Rev. Lett. 
{\bf 85}, 3520 (2000). 

\bibitem{tombesi} D. Vitali and P. Tombesi, Phys. Rev. A {\bf 59},
4178 (1999); e-print {\tt quant-ph/0108007}.

\bibitem{beth} P. Wocjan, M. R\"{o}tteler, D. Janzing, and T. Beth,
e-print {\tt quant-ph/0109063}; {\tt quant-ph/0109088}

\bibitem{lloyd} S. Lloyd and L. Viola,  e-print {\tt quant-ph/0008101},
Phys. Rev. A, Rapid Communication, in press.

\bibitem{wu} L.-A. Wu and D. A. Lidar, e-print {\tt quant-ph/01030039};
D. A. Lidar and L.-A. Wu, e-print {\tt quant-ph/0109021}.

\bibitem{dfs} E. M. Fortunato, L. Viola, J. Hodges, G. Teklemariam,
and D. G. Cory, ``Implementation of Universal Control on a 
Decoherence-Free Qubit'', submitted.

\bibitem{zanardi2} P. Zanardi,  Phys. Rev. A {\bf 63}, 012301 
(2001).

\bibitem{byrd} M. S. Byrd and D. A. Lidar, e-print 
{\tt quant-ph/0110121}.

\bibitem{carr} H. Y. Carr and E. M. Purcell, Phys. Rev. {\bf 94},
630 (1954).

\bibitem{mahler} J. A. Jones and E. Knill, J. Magn. Res. 
{\bf 141}, 322 (1999); D. W. Leung, I. L. Chuang, F. Yamaguchi, 
and Y. Yamamoto, Phys. Rev A {\bf 61}, 042310 (2000);
M. Stollsteimer and G. Mahler, e-print 
{\tt quant-ph/0107059}.

\bibitem{seth} S. Lloyd, Phys. Rev. Lett. {\bf 75}, 346 (1995).
 
\bibitem{preskill} J. Preskill, Proc. R. Soc. London A {\bf 454}, 
385 (1998), and references therein.

\bibitem{dfstheory} P. Zanardi and M. Rasetti, 
{ Phys. Rev. Lett.} {\bf 79} 3306 (1997); L.-M. Duan and G.-C. Guo, 
{\it ibid.} 1953 (1997); D. A. Lidar, I. L. Chuang, and K. B. 
Whaley, {\it ibid.} {\bf 81} 2594 (1998). 

\bibitem{nstheory} E. Knill, R. Laflamme, and L. Viola,
{Phys. Rev. Lett.} {\bf 84} 2525 (2000).

\bibitem{qubit} L. Viola, E. Knill, and R. Laflamme, 
{ J. Phys. A} {\bf 34} 7067 (2001).

\bibitem{eu}  D.P. DiVincenzo, D. Bacon, J. Kempe, G. Burkard, and 
K. B. Whaley, Nature {\bf 408}, 339  (2000); D. Bacon,  J. Kempe, 
D.P. DiVincenzo, D. A. Lidar, and K. B. Whaley, e-print 
{\tt quant-ph/0102140}.

\bibitem{kempe1} D. Bacon, J. Kempe, D. A. Lidar, and K.B. Whaley,
Phys. Rev. Lett. {\bf 85}, 1758 (2000).

\bibitem{note} Allowing for this degree of generality may be 
essential to succeeding in the explicit construction of desired 
encoded gates. For instance, when two DFS-qubits encoded into 
nearby clusters of 4 physical qubits are considered under 
collective noise, it was shown in \cite{kempe1} that an encoded 
controlled-phase gate can be constructed from a sequence of 
control operations which may map the computational space 
$\cH_{L_1}\otimes\cH_{L_2}$ into a larger irreducible subspace 
of $\cA'$ in the state space of 8 qubits, and back.  
See also \cite{qubit} for related observations.

\bibitem{ns} L. Viola, E. M. Fortunato, M. A. Pravia, E. Knill,
R. Laflamme, and D. G. Cory, Science {\bf 293}, 2059 (2001).  

\bibitem{kempe} J. Kempe, D. Bacon, D. A. Lidar, and K. B. 
Whaley, { Phys. Rev. A} {\bf 63} 042307 (2001).

\bibitem{composite} R. Freeman, S. P. Kempsell, and M. H. Levitt, 
J. Magn. Res. {\bf 38}, 453 (1980); M. H. Levitt, {\it ibid.} 
{\bf 48}, 234 (1982). 

\bibitem{fastpulse} E. M. Fortunato, M. A. Pravia, N. Boulant, 
G. Teklemariam, T. F. Havel, and D. G. Cory, ``Design of Strongly
Modulating Pulses to Implement Precise Effective Hamiltonians 
for Quantum Information Processing'', submitted.

\bibitem{concat} D. A. Lidar, D. Bacon, and K. B. Whaley, 
{Phys. Rev. Lett.} {\bf 82}, 4556 (1999). 


\bibitem{leggett} A. J. Leggett {\it et al.}, Rev. Mod. Phys. 
{\bf 59}, 1 (1987).

\bibitem{bacon} D. A. Lidar, D. Bacon, J. Kempe, and K. B. 
Whaley, { Phys. Rev. A } {\bf 63} 022306 (2001).

\end{thebibliography}
\end{document}